\documentclass[%
 reprint,
superscriptaddress,
amsmath,amssymb,
aps,
prb,
floatfix,
]{revtex4-2}
\usepackage{float}
\usepackage{natbib}
\usepackage{graphicx}
\usepackage{dcolumn}
\usepackage{bm}
\usepackage[dvipsnames]{xcolor}
\usepackage[colorlinks=true]{hyperref}
\hypersetup{
    colorlinks=true,
    linkcolor=blue,
    citecolor=blue,
    urlcolor=blue
}

\begin{document}

\preprint{APS/123-QED}

\title{Reconfigurable Filamentary Conduction in Thermally Stable 
  Zeolitic Imidazolate Framework (ZIF-8) Resistive Switching Devices}

\author{Divya Kaushik}%
\affiliation{CSIR-National Physical Laboratory, Dr. KS Krishnan Marg, New Delhi-110012, India.}
\affiliation{Academy of Scientific and Innovative Research (AcSIR), Ghaziabad-201002, India.}
\author{Nitin Kumar}%
\affiliation{Department of Physics, University at Buffalo-SUNY, 239 Fronzack Hall, Buffalo, NY 14260 US}
\author{Harshit Sharma}%
\affiliation{CSIR-National Physical Laboratory, Dr. KS Krishnan Marg, New Delhi-110012, India.}
\affiliation{Academy of Scientific and Innovative Research (AcSIR), Ghaziabad-201002, India.}
\author{Pukhraj Prajapat}%
\affiliation{CSIR-National Physical Laboratory, Dr. KS Krishnan Marg, New Delhi-110012, India.}
\affiliation{Academy of Scientific and Innovative Research (AcSIR), Ghaziabad-201002, India.}
\author{Mehamalini V.}%
\affiliation{Department of Nanotechnology, Amity University, Noida, Uttar Pradesh, India.
}
\author{G.Sambandamurthy}
\email{sg82@buffalo.edu}
\affiliation{Department of Physics, University at Buffalo-SUNY, 239 Fronzack Hall, Buffalo, NY 14260 US}
\author{Ritu Srivastava}
\email{ ritu@nplindia.org}
\affiliation{CSIR-National Physical Laboratory, Dr. KS Krishnan Marg, New Delhi-110012, India.}
\altaffiliation{Academy of Scientific and Innovative Research (AcSIR), Ghaziabad-201002, India.}

\begin{abstract}

The rapid growth of digital technology has driven the need for efficient storage solutions, positioning memristors as promising candidates for next-generation non-volatile memory (NVM) due to their superior electrical properties. Organic and inorganic materials each offer distinct advantages for resistive switching (RS) performance, while hybrid materials like metal-organic frameworks (MOFs) combine the strengths of both. In this study, we present a resistive random-access memory (ReRAM) device utilizing zeolitic imidazolate framework (ZIF-8), a MOF material, as the resistive switching layer. The ZIF-8 film was synthesized via a simple solution process method at room temperature and subsequently characterized. The Al/ZIF-8/ITO device demonstrates bipolar resistive switching behaviour with an on/off resistance ratio of $\sim10^2$, stable retention up to $10^4$ seconds, and consistent performance across 60 cycles while exhibiting robust thermal stability from $-20^\circ\mathrm{C}$ to $100^\circ\mathrm{C}$. Low-frequency noise and impedance spectroscopy measurements suggest a filamentary switching mechanism. Additionally, the memory state can be tuned by adjusting the reset voltage, pointing to potential as multi-level memory. Potentiation and depression experiments further highlight the device’s promise for neuromorphic applications. With high stability, tunability, and strong performance, the ZIF-8 based ReRAM shows great promise for advanced NVM and neuromorphic computing applications.

\end{abstract}
\maketitle
\section{Introduction}
The growing demand for advanced digital memory in the AI-driven era is spurring innovation to address challenges such as scalability limits, stability issues, slower speeds, higher power consumption, and lower bit density.\cite{Yan2022} Traditional devices, including  dynamic random access memory (DRAM), static random access memory (SRAM), and flash memory, suffer from these limitations, and even complementary metal-oxide-semiconductor (CMOS) technology itself faces challenges in terms of scalability. An ideal memory device should combine high endurance and retention, low power consumption, nanosecond operation, and enhanced scalability for greater bit density, making memristors a focal point of extensive recent research.\cite{Ielmini2018, DUAN2024} Memristors, also known as resistive random-access memories (ReRAMs), offer several advantages over conventional memory devices, including lower power consumption, faster operation, higher on/off ratios, and structural as well as fabrication simplicity.\cite{Ding2023_SS} A memristor structure typically comprises two electrodes—one on the top and the other at the bottom—with a dielectric or insulating layer sandwiched between them, forming a metal-insulator-metal (MIM) configuration that enables its function as a memory device.\cite{Chen2023} Metal-oxide-based insulating materials are often used as the active layer in memristors, while the electrodes can vary, typically consisting of elemental metals, metal alloys, nitrides, or doped conductive materials.\cite{Wang2020,Ding2019} Conduction in memristors occurs through the formation and disruption of conductive filaments, with mechanisms for filament formation including charge hopping, charge trapping and de-trapping, redox reactions, interfacial interactions, ion migration, and thermochemical processes.\cite{Chen2023,Ding2023,Gao2019} The demand for active layer materials has led to the development of alternatives to metal oxides, such as porous and hybrid materials like metal-organic frameworks (MOFs), which are easier to synthesize and hold significant potential for memory and neuromorphic computing applications.\cite{Ding2023,Xu2024,Shu2022}\\

MOF-based memristors have garnered significant attention due to their memristive properties and ease of synthesis. These materials are porous, crystalline structures known for their thermal and chemical stability, as well as their luminescent characteristics. However, the exploration of these materials for memristor applications remains limited.\cite{Kitagawa2014, Yusuf2022} These micro- to mesoporous materials offer tunable porosity, allowing for customization, which makes MOFs highly functional material for various applications.\cite{Rowsell2004, Bull2022} Efficient charge transport within memristors can be enhanced by utilizing porous structures. By tuning synthesis parameters such as precursor concentration and temperature, the pore size of MOFs can be controlled, enabling the accommodation of guest molecules smaller than the pore size, which in turn improves their performance and functionality.\cite{Zhou2023} Various MOF-based memristors, including UiO-66, HKUST-1, ZIF-67, and MIL-53, have been extensively studied.\cite{Yi2019, Pan2015, Kaushik2024, Xu2024}

Zeolitic-Imidazolate frameworks (ZIFs), as a subset of MOFs, are zeolite-like frameworks with metal ions interconnected by imidazolate ligands, which are promising highly porous materials for diverse applications. ZIF-8, a type of ZIFs, resembles a zeolitic structure but contains zinc metal ions and 2-methyl imidazolate (2-MIm) organic linkers.\cite{Zheng2023} It has a sod (sodalite) structure similar to that of zeolite. Functionalization of zeolites is challenging but ZIF-8 can be modified. ZIF-8 also has various advantages like their stability at even 550$^{\circ}$C and mechanical stability that they can even withstand a 28GPa pressure and still they can revert back to their original state. Thermogravimetric analysis (TGA) also indicated that ZIF-8 did not exhibit any weight loss up to 150$^{\circ}$C. Even after dissolution in various solvents, ZIF-8 seems to be stable in terms of its crystallinity and porosity.\cite{Pouramini2023} Their insulating behavior and higher chemical and thermal stability make them a great candidate for use in environments of higher temperatures or pressures. ZIF-8 has various applications such as micro and nanofiltration, forward and reverse osmosis, electrochemical sensors, batteries, resistive switching, catalysis, and heavy metal removal from wastewater, and can act as a host for various biological constituents such as proteins and enzymes, gas separation and absorption, and drug delivery.\cite{Bergaoui2021,Chen2024,Eddaoudi2000,Mohan2023} 

The ZIF-8 based MIM device can be used for digital or analog applications based on the switching speed and SET/RESET voltages. Notably, ZIF-8-based memristors exhibit both multilevel resistive switching, showcasing their versatility for advanced memory applications, and excellent electrical stability, making them reliable for long-term use.\cite{Liu2016} The delocalized $\pi$ bond in the ZIF-8 structure helps in charge transport and conductive filament formation. They contain zinc ions that are tetrahedrally connected to four 2-MIm linkers (ZIF-8 structure is shown in Figure \ref{Fig :figure1} (a)). \cite{Bergaoui2021} Zinc is connected to the nitrogen group of MIm; hence, it is ditopic.\cite{Yusuf2022} Owing to their porous channels, ions or charges can be entrapped within and contribute to the formation of conductive paths. However, the conduction mechanism in memristive devices remains unclear, which is crucial for parameter optimization and the practical use of these devices.

In previous studies, efforts have been made to enhance resistive switching and understand the underlying conduction mechanism in MOF-based memristors.\cite{Liu2016,Jeon2021,Park2017,He2022} However, information reported on low-frequency noise spectroscopy in ZIF-8 memristor devices is very limited, despite the fact that current fluctuations during read operations can lead to significant errors in determining memory states .\cite{Spinelli2008} Low-frequency noise spectroscopy has proven to be a powerful tool for investigating switching mechanisms in various types of resistive memories.\cite{Lee2010, Zhao2014} The power spectral density of current fluctuations typically follow a power law of \(1/f^{\alpha}\) with \(\alpha \approx 1\).\cite{Ali2017} Fluctuations where \(\alpha \approx 1.5\) suggest different physical origins than \(\alpha \approx 1\), often attributed to long-range diffusion processes.\cite{Kar2002, Nevins1990, Scofield1985} Additionally, a deeper understanding of the switching mechanisms can be gained by performing low-frequency noise spectroscopy at voltage-driven resistive switching and across different conduction regimes.\cite{Carbone2005,Hu2021}

In this study, ZIF-8 was synthesized through a simple solution-based precipitation method that avoids high-temperature requirements. Metal-insulator-metal structure was fabricated using an Indium Tin Oxide (ITO)-coated glass substrate as the bottom electrode and aluminum (Al) as the top electrode with ZIF-8 as active layer. The resulting Al/ZIF-8/ITO device demonstrated stable bipolar resistive switching with excellent reproducibility and thermal stability. The conduction mechanism revealed that the device initially exhibits ohmic conduction in the high-resistance state (HRS), transitions to Schottky emission with increasing voltage, and further voltage increase leads to space charge-limited conduction before returning to ohmic behavior in the low-resistance state (LRS). Impedance spectroscopy suggests filament formation during the transition from HRS to LRS. In ultra-low frequency noise measurements the increase in power spectral density around the transition attributed to Zn ion movement forming conductive filaments. Noise measurements indicate no significant ion movement during decreasing voltage, suggesting the filaments remain intact until a negative bias is applied. Notably, the filament length can be adjusted by varying the reset voltage, enabling the ZIF-8 device to support multi-level memory applications in addition to its functionality as ReRAM and memristors.  

\section{Experimental Section}
\subsection{Materials }
The materials used in this experiment were obtained from Sigma-Aldrich and used as obtained without further purification. The chemicals used are Zinc (II) nitrate hexahydrate (Zn(NO$_3$)$_2$·6H$_2$O, ACS reagent, $\geq$ 98 $\%$), 2-Methylimidazole (C$_4$H$_6$N$_2$, 99$\%$) and Methanol (CH$_3$OH, ACS reagent, $\geq$99.8$\%$).

\subsection{Synthesis of ZIF-8 }
A 0.45g of zinc nitrate is taken in 1.5mL of methanol and allowed to stir for 10 minutes at 500 rpm under room temperature. Similarly, a 2-MIm solution is made using 5.5g of 2-methyl Imidazolate in 20mL of methanol and stirred under similar conditions. After the constituents are completely mixed, add the 2-MIm solution dropwise to the Zinc nitrate solution till it turns white. It is then allowed to stir for another 24 hours at 500 rpm under room temperature. Centrifuge it for 5 mins at 6000 rpm and then it washed with methanol several times. ZIF-8 is hence synthesized. Store it in methanol for further use.

\subsection{Device fabrication }

The ITO-coated glass substrates were cut into 15mm × 15mm sections and cleaned with soap solution and then ultrasonicated with DI water, acetone and 2-propanol for 15 min each respectively. The cleaned substrate were dried in vacuum oven. On the cleaned dried ITO-coated glass substrates, ZIF-8 was spin coated at 2000rpm for 30 sec with annealing time of 5min at 120°C which provides a $\sim$150nm thick ZIF-8 active layer. The thickness of ZIF-8 is measured by stylus profilometer. Then a 100nm thick top aluminum electrode was deposited via thermal evaporation under a vacuum pressure of 5 × 10$^{-6}$ mbar. A shadow mask was used during deposition to create a top electrode with dimensions of 200$\times$200$\mu$m$^2$.

\subsection{Electrical Transport and Characterization}
The electrical transport properties were examined using Rucker and Kolls Model 260 or SUSS MicroTec probe stations. Current-voltage measurements were conducted with a Keithley 2450 Source Meter Unit (SMU), utilizing Test Script Processor (TSP) commands for precise control.

\section{Results and discussion}

\begin{figure*}[!h]
      \centering
      \includegraphics[width=\textwidth]{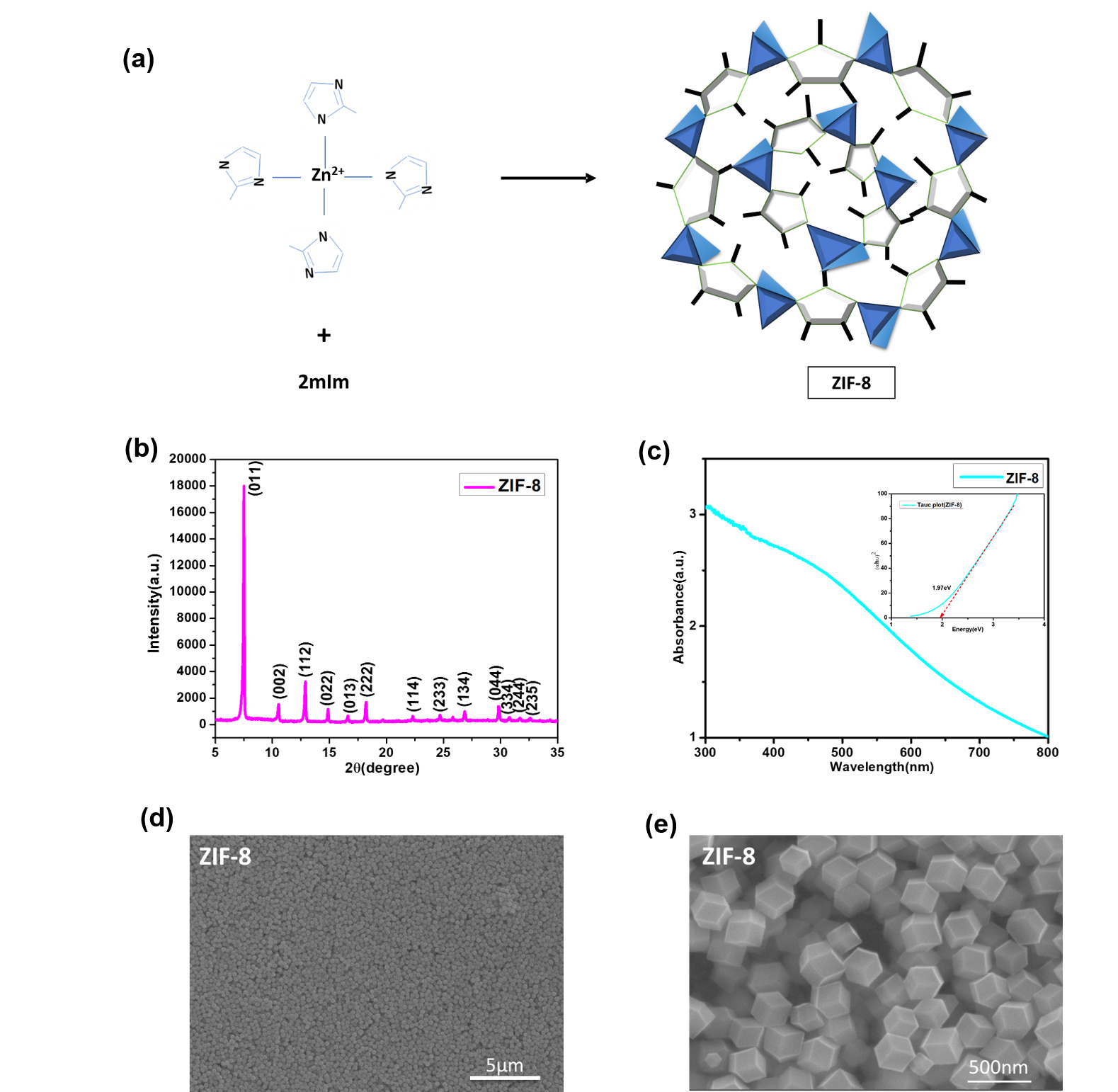}
      \caption
      { (a) Schematic diagram of formation of the Three-dimensional structure of ZIF-8 (b) X-Ray diffraction pattern of ZIF-8 (c)  UV-visible absorbance behavior of ZIF-8 (inset figure: corresponding tauc plot of ZIF-8)  (d) FESEM micrograph of ZIF-8 thin film on ITO substrate (e) zoom FESEM micrograph of ZIF-8 thin film on ITO substrate.}
      \label{Fig :figure1}
\end{figure*}

XRD spectroscopy was employed in analysis the crystallographic structure of ZIF-8, using a Cu-K alpha line of 0.1541nm with angle ranging from 5$^{\circ} $-40$^{\circ} $. The ZIF-8 which has been synthesized demonstrated a cubic phase and excellent crystallinity.\cite{Paul2018} It has the crystallographic planes (011),(002),(112),(022),(013),
(222), (114), (233), (134), (044), (334), (244) and (235) as shown in Figure \ref{Fig :figure1} (b)\cite{Zhu2012} which agrees with the JCPDS file 00-062-1030 of the reported ZIF-8. The UV Visible absorption spectroscopy also studied in which the  absorption is identified in UV to Visible spectrum as indicated in the fig1(c), furthermore the Tauc Plot for the same is represented in the inset of the Figure \ref{Fig :figure1} (c) showing a band gap of 1.97 eV. The morphological analysis of ZIF-8 has been done using Field Emission Scanning Electron Microscopy (FESEM). It reveals that ZIF-8 has a rhombic dodecahedron structure.\cite{Lee2015,Sun2016} The particle size is homogeneous and uniformly distributed over the substrate as shown in Figure \ref{Fig :figure1} (d). It reveals that ZIF-8 has a tetrahedral structures as shown in Figure \ref{Fig :figure1} (e).  

\begin{figure*}[t]
\centering
\includegraphics[width=\textwidth]{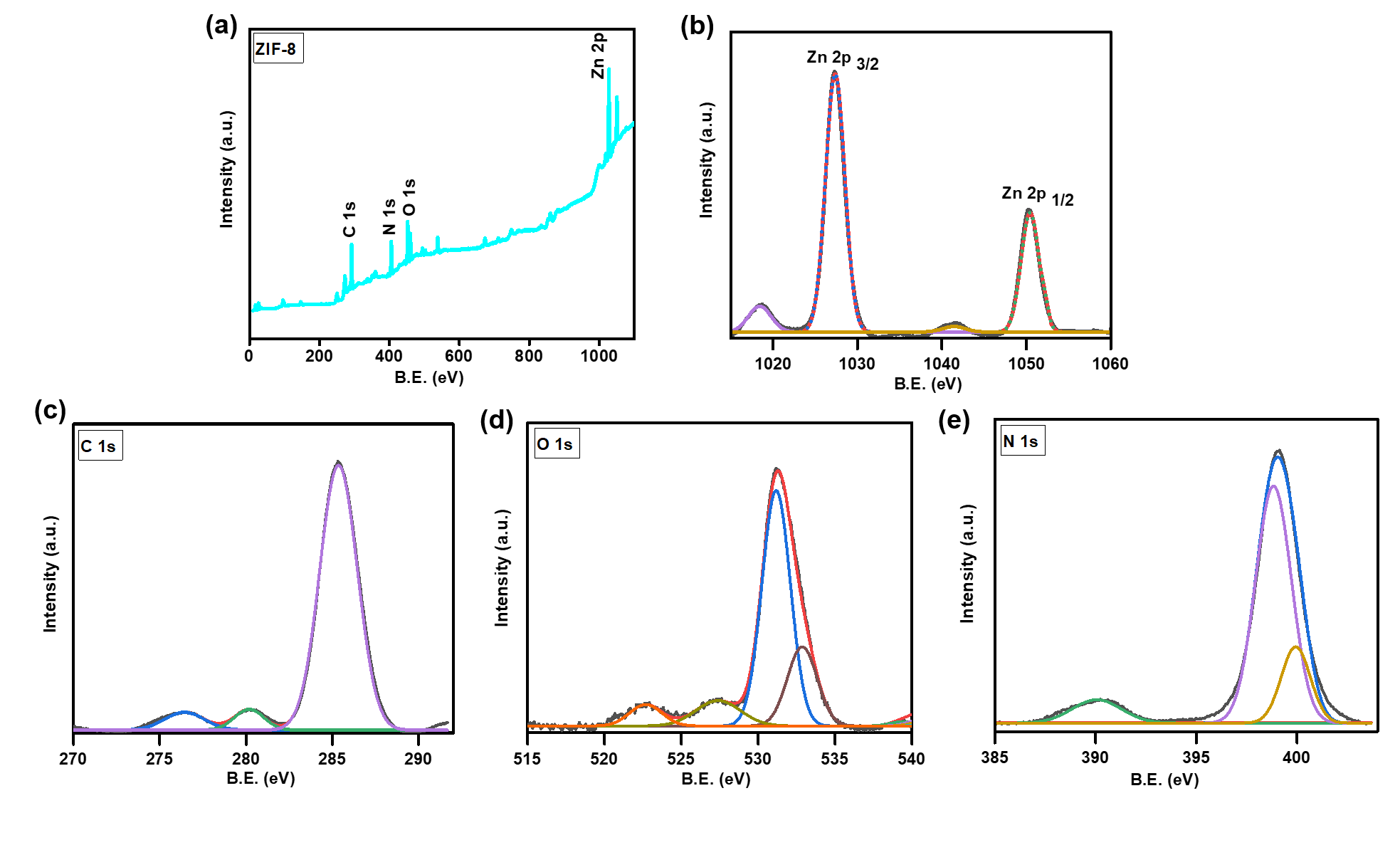}
\caption{XPS analysis spectra of the ZIF-8 (a) Survey scan and (b),(c),(d),(e) the core level spectra of Zn 2p, N 1s, C 1s, and O 1s respectively.}
\label{fig:2}
\end{figure*}

The X-ray Photoelectron Spectroscopy (XPS) analysis of ZIF-8 is presented in Figure \ref{fig:2}. The survey spectrum and high-resolution spectra for C 1s, O 1s, and N 1s provide insights into the surface composition and bonding environments of the material. Figure~\ref{fig:2} (a) shows the wide-scan survey spectrum, which reveals distinct peaks corresponding to C, N, and O, confirming the primary components of the ZIF-8 framework. Minor oxygen peaks are attributed to adsorbed species such as H$_2$O or CO$_2$, typical in porous metal-organic frameworks (MOFs) exposed to ambient air.\cite{Li2021} In Figure~\ref{fig:2} (b), the core level spectra of Zn 2p exhibit two peaks that represent Zn are adjacent to Zn 2p$^{1/2}$ and Zn 2p$^{3/2}$. In Figure~\ref{fig:2} (c), the high-resolution C 1s spectrum shows a dominant peak at 284.8 eV, which corresponds to C–C and C–H bonds from the organic linker in ZIF-8. A smaller peak at approximately 286 eV is attributed to C–N bonding, confirming the imidazolate linker’s presence.\cite{Tian2014} The O 1s spectrum in Figure~\ref{fig:2} (d) shows a peak around 532 eV, which can be assigned to oxygen species adsorbed on the surface, such as O–H and possibly C=O, commonly observed in MOFs after exposure to the environment.\cite{Wahab2007} Lastly, Figure~\ref{fig:2} (e) presents the N 1s spectrum, with a prominent peak at 398 eV, corresponding to nitrogen in the imidazolate ring. Deconvolution suggests slight variations in the nitrogen bonding environment, in agreement with previous reports on nitrogen bonding in MOF structures.\cite{Sandomierski2022} These XPS results confirm the integrity of the ZIF-8 structure, with the expected chemical states of C, N, and O present, and provide essential insights for its application in resistive switching devices.

\begin{figure*}[t]
\centering
\includegraphics[width=\textwidth]{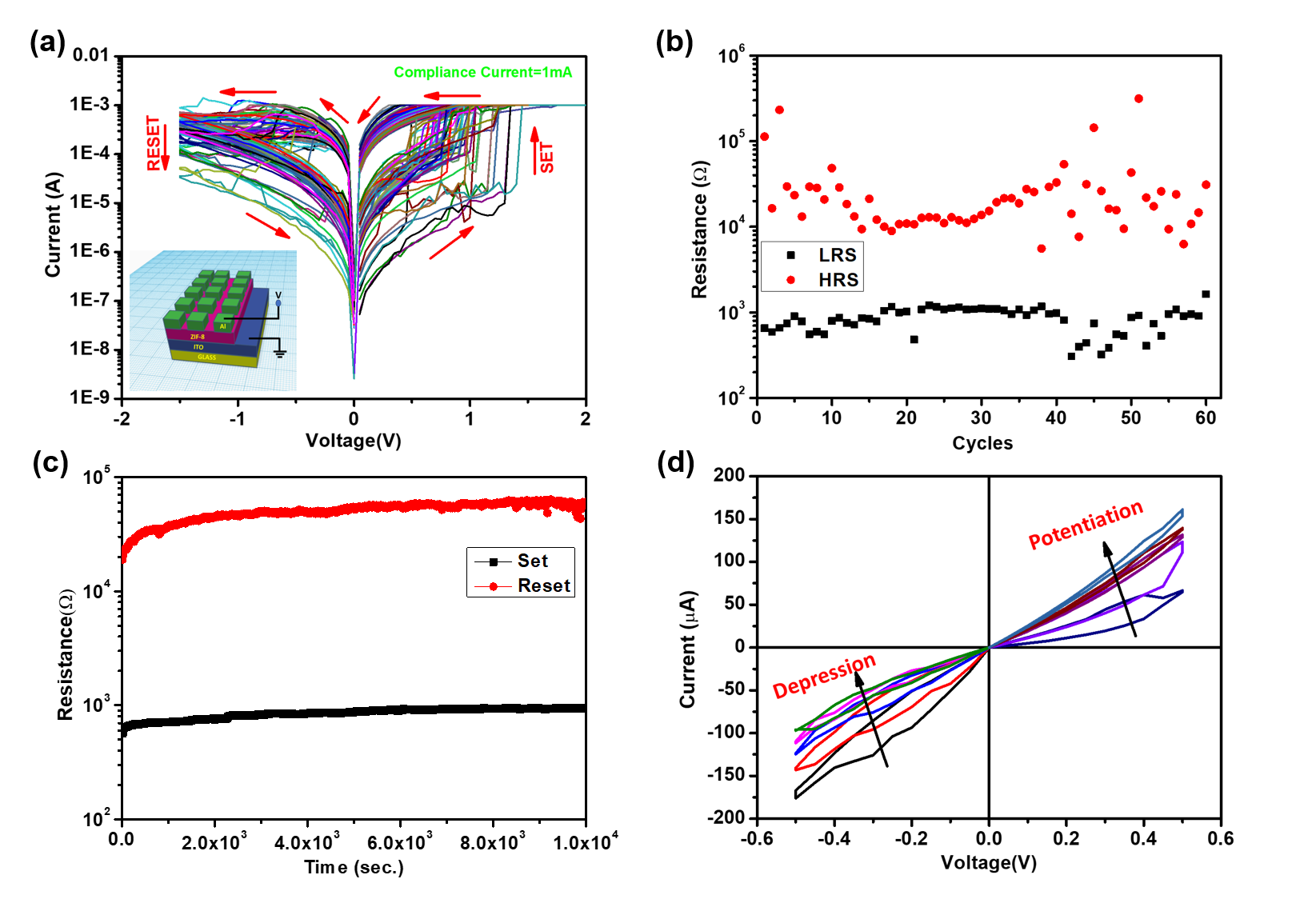}
\caption {(a) I-V characteristics of Al/ZIF-8/ITO device showing the 60 different cycles of bipolar resistive switching (inset figure: Schematic diagram of the Al/ZIF-8/ITO device), (b) Endurance of the device for 60 cycles, (c) Retention of the LRS and HRS with time, and (d) Potentiation and Depression for the 5 different cycles. }
\label{fig:Figure3}
\end{figure*}

The fabricated schematic device structure Al/ZIF-8/ITO (inset of Figure~\ref{fig:Figure3} (a)) is studied for its basic resistive switching behavior. A cyclic voltage of 0V $\rightarrow$ 2V $\rightarrow$ 0V was applied to the device, while keeping the compliance current (CC) at 1 mA to protect the device. Initially, the device was in HRS state but before reaching the 2V the device current showed a sharp jump in current to 1 mA (CC), showing a transition from HRS to LRS. The device remains in LRS for 2V $\rightarrow$ 0V sweep exhibiting non-voltaile memory swithcing. To reset the device back into the HRS, a voltage sweep cycle of 0V $\rightarrow$ -2V $\rightarrow$ 0V was applied, during 0V $\rightarrow$ -2V sweep device initially remains in the LRS but at some particular voltage device current decreases rapidly and the device switches back into the HRS. For further -2V $\rightarrow$ 0V sweep device remains in HRS. The variation in the switching voltages for 60 different cycles have been shown in Figure~\ref{fig:Figure3} (a). The endurance cycle lifetime of the device has also been studied upto 60 cycles of Set/Reset measurements as shown in Figure~\ref{fig:Figure3} (b). More variations have been observed in HRS resistances than for the LRS because the reset process is more complex than the set process. The retention at room temperature ($\sim$ 25°C) of the LRS and HRS have been also studied as shown in Figure~\ref{fig:Figure3} (c). For this, the device has been set in LRS and HRS  and then the resistance of the device has been read by applying a read voltage of 100mV. The device shows a high retention up to 10$^4$ seconds and a high HRS to LRS resistance ratio of the order of $\sim$10$^2$ has also been observed.
Analog switching characteristics have also been observed in Al/ZIF-8/ITO devices.\cite{Jeon2021} The conductance of the Al/ZIF-8/ITO memristor demonstrates behavior similar to synaptic weight modulation, showing both potentiation and depression characteristics.\cite{Wu2023} The device conductance evolves gradually with voltage sweeps, mimicking biological memory and forgetting processes through potentiation and depression behaviors, respectively.\cite{Mannion2023} Figure~\ref{fig:Figure3} (d) clearly illustrates the gradual and repeatable conductance modulation of the Al/ZIF-8/ITO device.  For potentiation, successive voltage sweep cycles from 0 to 0.6 V were applied, while for depression, consecutive sweep cycles from 0 to -0.6 V were applied. With positive voltage sweeps from 0 to 0.6 V, the current progressively increases with each cycle, indicating potentiation in conduction. This behavior mirrors a memorization process, as the device "learns" by gradually strengthening its conductive pathways. In contrast, applying negative voltage sweeps from 0 to -0.6 V gradually decreases the current with progressive sweep cycles, indicating conductance depression. This behavior corresponds to the "forgetting" process, as the conductive pathways weaken over successive cycles. This gradual change in conductance highlights the device's analog switching characteristics, crucial for emulating synaptic plasticity.\cite{Shu2022} Such continuous modulation of conductance is key to neuromorphic applications, where a memristor’s ability to mimic synaptic functionality could enable energy-efficient, brain-like computing.\cite{Hu2018,Subin2021}

To evaluate the reproducibility of the RS behavior in Al/ZIF-8/ITO devices, seven different devices from three separately fabricated samples, all processed under identical conditions, were tested. Figure S1(a)-(g) displays the RS behavior of these devices, each exhibiting consistent bipolar switching with minimal variation in switching voltages. The near-identical I-V characteristics across these devices underscore the reproducibility and uniformity of the Al/ZIF-8/ITO structure. Figure S1(h) illustrates the LRS and HRS values for the same seven devices, where both states exhibit a consistent resistance distribution with only minor fluctuations. This uniformity across devices highlights the high device-to-device consistency of the bipolar resistive switching behavior, further validating the reproducibility of the fabricated devices. In addition to device uniformity, the temperature-dependent RS behavior of the Al/ZIF-8/ITO device was also investigated, with performance evaluated over a wide temperature range from -20°C to 100°C. Figure S2(a)-(m) shows the I-V characteristics of the device at different temperatures, illustrating consistent bipolar RS behavior across this temperature range. The data indicate a decreasing trend in HRS resistance as temperature increases, while LRS resistance varies randomly, as shown in Figure S2(m). The reduction in HRS resistance with temperature is due to the presence of thermally generated charge carriers at higher temperatures. Meanwhile, the LRS resistance shows random behavior due to the inherently variable nature of the set process. Although the HRS-to-LRS resistance ratio decreases with temperature, distinct switching states with sufficient resolution are observed across a wide range (-20°C to 100°C). This indicates reliable device performance under various thermal conditions\cite{Kumari2021,Sato2023}, reinforcing its potential for practical in-memory applications where thermal stability is critical.\cite{Pazos2024} This manuscript focus transitions from discussing the device's properties, performance and applications to exploring the conduction mechanism, crucial for optimizing its parameters for practical use.

\begin{figure*}[!ht]
      \centering
      \includegraphics[width=0.9\textwidth]{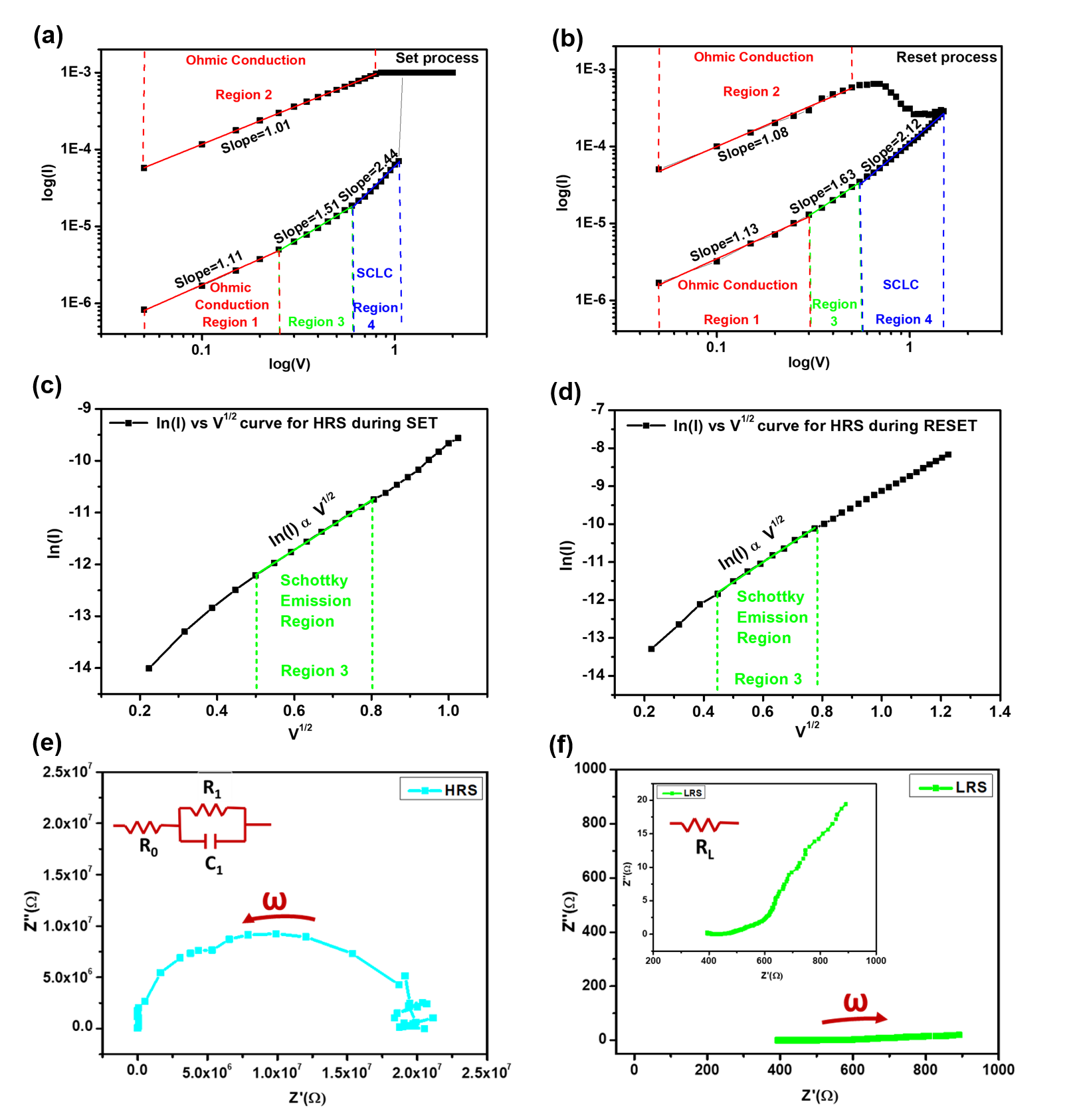}
      \caption
      { (a) Logarithmic plot of I-V curves of Set process and (b) Reset process with linear regression, and ln(I) vs V$^{1/2}$ curves for the HRS of the device during (c) SET, and (d) RESET process, (e) cole-cole plot for HRS and (f) LRS (inset figure: zoomed out of LRS). }
\label{fig:Figure4}
      
\end{figure*}

To examine the conduction mechanism in the Al/ZIF-8/ITO device current-voltage data for the SET and RESET processes were fitted to different electron transport models. The I-V characteristics for both processes were initially plotted on a Log${10}$(I) - Log${10}$(V) scale as shown in Figure \ref{fig:Figure4} (a) and Figure \ref{fig:Figure4} (b). Through linear regression analysis, three distinct regions (1, 3, and 4) were identified in HRS with slopes of 1.11, 1.51, and 2.44 for the SET process, and 1.13, 1.63, and 2.12 for the RESET process. In LRS, only one region (region 2) was observed, with slopes of 1.01 for the SET process and 1.08 for the RESET process. Regions 1 and 2, with slopes near 1 (I $\propto$ V), were identified as regions of ohmic conduction. In region 4, where the slope is approximately 2 (I $\propto$ V$^2$), space-charge-limited conduction (SCLC) was observed. The slope of region 3, approximately 1.5, did not correspond to either ohmic or SCLC behavior. To investigate this, a Schottky-emission (S-E) mechanism was considered where ln(I) should be directly proportional to $\sqrt{V}$. To verify this, ln(I) was plotted against $\sqrt{V}$ for the HRS in both SET and RESET processes (as shown in Figure~\ref{fig:Figure4} (c) and (d)). The linear fit in region 3 confirmed that ln(I) $\propto$ $\sqrt{V}$, identifying this as the S-E region in the I-V characteristics. The current density expressions for these three distinct conduction mechanisms namely ohmic conduction, space charge limited conduction, and Schottky-emission conduction  are as follows\cite{Lim2015,Li2011}: 

\textbf{Equations:}

\begin{equation}
    J_{\text{OC}} = \sigma E = q \mu N_C E \exp\left( \frac{-(E_C - E_F)}{kT} \right)
\end{equation}
\begin{equation}
    J_{\text{SCLC}} = \frac{9}{8} \varepsilon_i \mu \theta \frac{V^2}{d^3}
\end{equation}
\begin{equation}
    J_{\text{SE}} = \frac{4\pi q m^* (kT)^2}{h^3} \exp\left( \frac{-q\left(\Phi_B - \sqrt{\frac{qE}{4\pi \varepsilon}}\right)}{kT} \right)
\end{equation}

\textbf{Where:} \( J_{\text{SE}} \) is the current density for Schottky emission, \( J_{\text{ohmic}} \) is the current density for ohmic conduction, \( J_{\text{SCLC}} \) is the current density for Space-Charge-Limited Conduction (SCLC), \( q \) is the charge of an electron, \( m^* \) is the electron effective mass in the active layer, \( k \) is Boltzmann’s constant, \( T \) is the temperature, \( h \) is Planck’s constant, \( \Phi_B \) is the Schottky barrier height, \( E \) is the electric field, \( \varepsilon \) and \( \varepsilon_i \) are the permittivities of the active layer, \( \mu \) is the electron mobility, \( N_C \) is the effective density of states of the conduction band, \( E_C \) is the conduction band energy, \( E_F \) is the Fermi energy level, \( \theta \) is the ratio of free and shallow trapped charge, \( V \) is the applied voltage, \( d \) is the thickness of the active layer, and \( \sigma \) is the electrical conductivity.
The distinct regions observed in the I-V characteristics of the Al/ZIF-8/ITO device illustrate different conduction mechanisms activated by increasing voltage. At low voltages in the HRS, the device initially follows ohmic conduction, with a linear relationship between current and voltage (I $\propto$ V), reflecting direct and unobstructed charge transport through the material. After this initial ohmic region, as the voltage rises, the device enters a Schottky-emission (S-E) region. Here, the current shows a linear dependence of ln(I) on $\sqrt{V}$, consistent with thermionic emission over the Schottky barrier at the metal/semiconductor interface. In this region, the applied electric field assists in overcoming the barrier and facilitating carrier injection.\cite{Lim2015,JeonYoun2021} At even higher voltages, the device transitions into the SCLC region. In this regime, the current follows a V$^2$ dependence, indicating that injected carriers are building up in the material, creating a space-charge effect that enhances conduction beyond simple ohmic behavior. This increase in current is characteristic of space-charge injection, where the injected carriers outnumber those naturally present in the material. Following the SCLC region, the device enters the low-resistance state (LRS), where it once again follows ohmic conduction. This final ohmic behavior in the LRS indicates the establishment of a stable, low-resistance pathway within the material, allowing for linear, unobstructed current flow at this stage.

\begin{figure*}[t]
\centering
\includegraphics[width=\textwidth]{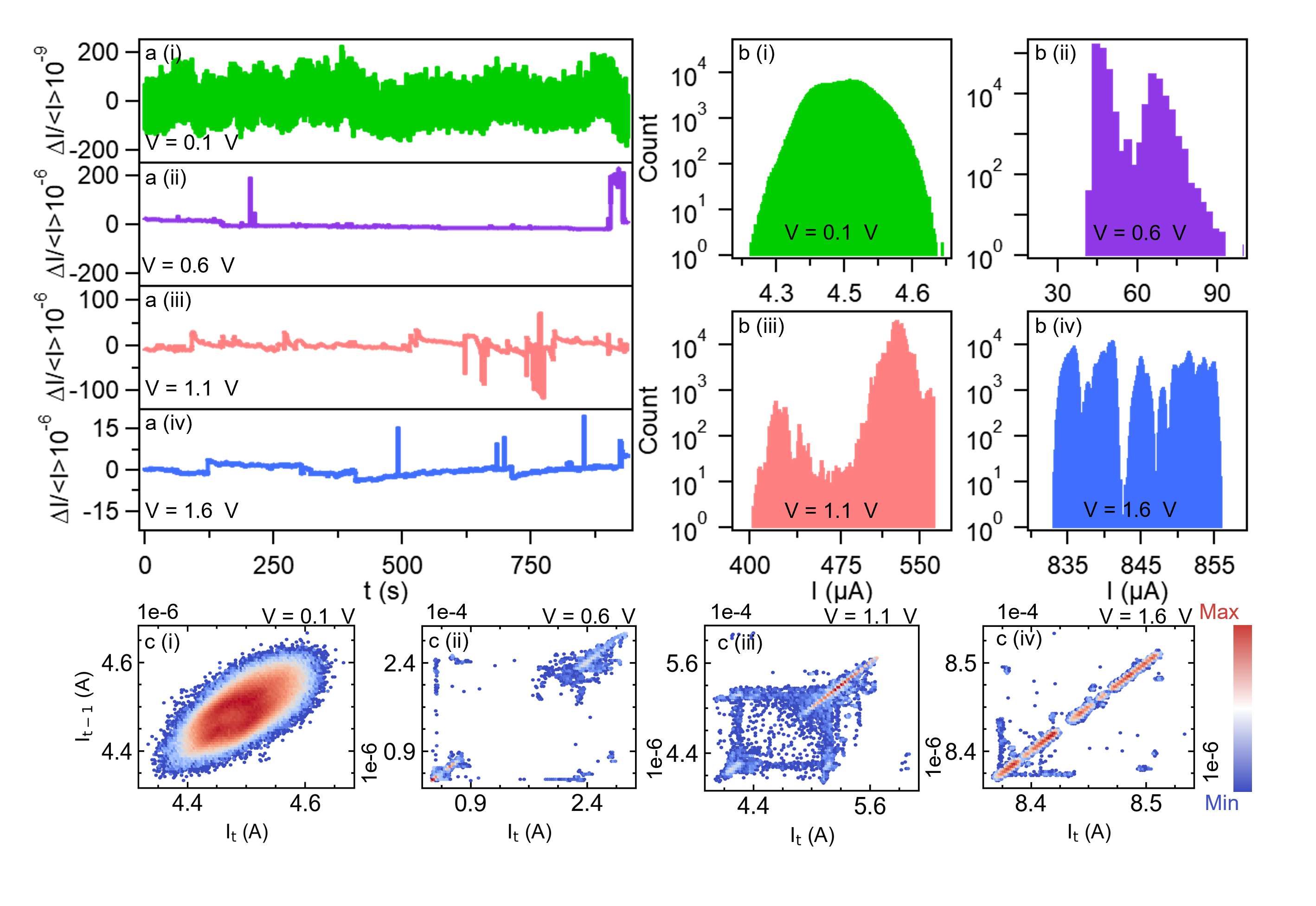}
\caption{Low-frequency noise analysis in time domain for the high resistance state (HRS) at four different voltages. (a) Residual current fluctuations as a function of time at (i) 0.1 V, (ii) 0.6 V, (iii) 1.1 V, and (iv) 1.6 V applied voltage, showing distinct behaviors in different conduction regions. (b) Probability distribution function (PDF) of the fluctuations for the time series in (a), showing Gaussian behavior at 0.1 V and deviation from Gaussian at higher voltages. (c) Time-lag plot of the fluctuations with a lag order of 1 for the time series in (a) highlighting the stable states and transitions.}
\label{fig:6}
\end{figure*}

\begin{figure*}[!ht]
\centering
\includegraphics[width=\textwidth]{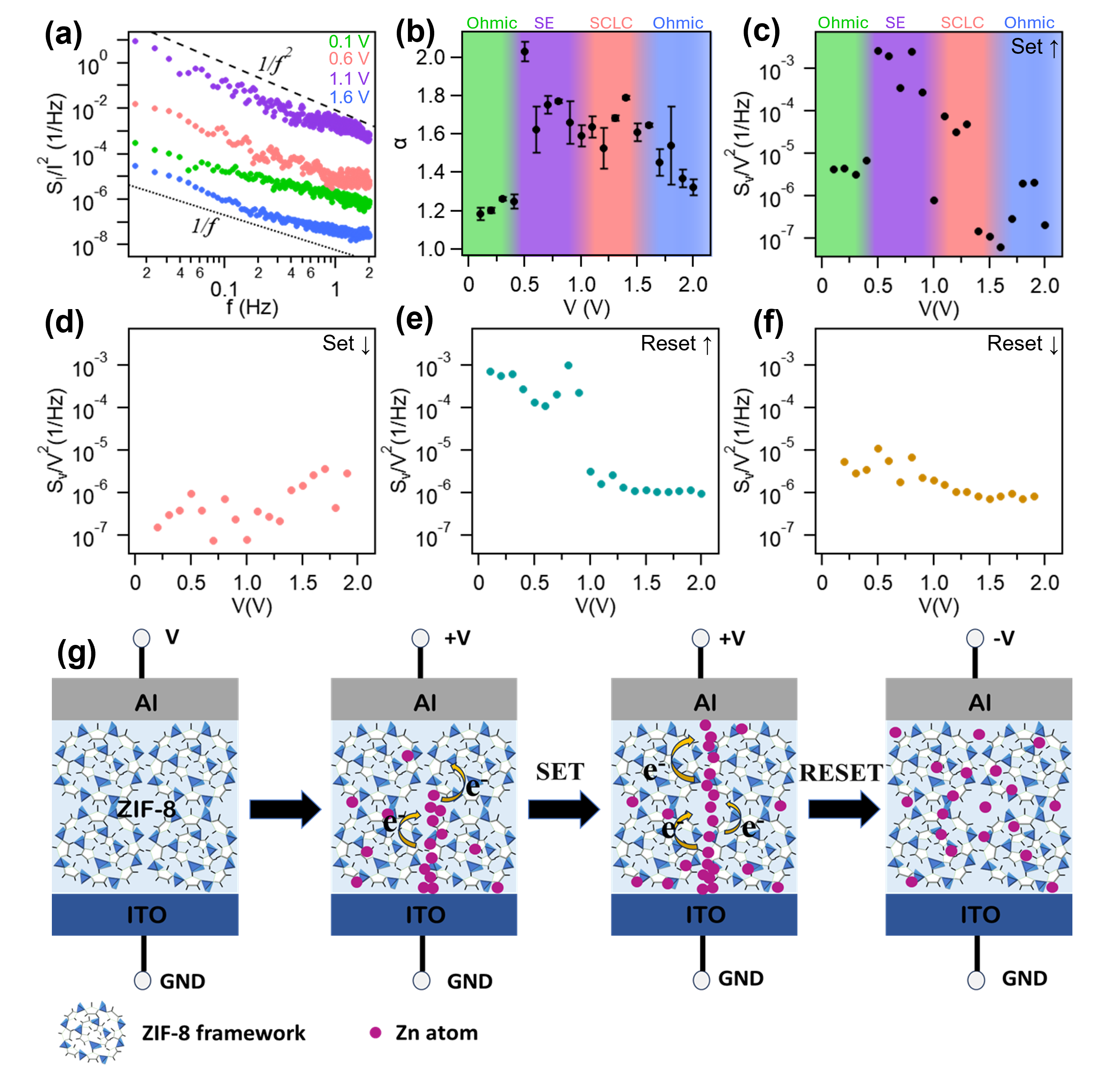}
\caption{Frequency-domain analysis of low-frequency noise during the set and reset operations of the device. (a) Normalized power spectral density (PSD) in the high-resistance state (HRS). The black lines represent \(1/f^{\alpha}\) scaling with exponents of 1 (dashed) and 2 (dotted), highlighting the changes in noise behavior with increasing voltage. (b) Frequency-dependent PSD slope (\(\alpha\)) across four distinct conduction regions during the Set~\(\uparrow\) cycle, with each region clearly delineated. The error bars represent the standard deviation from three different runs on the same device. (c) Noise magnitude as a function of applied voltage for the Set~\(\uparrow\) sweep, with the same conduction regions identified. (d) Noise magnitude for the Set~\(\downarrow\) sweep. (e, f) Noise magnitude trends during Reset~\(\uparrow\) and Reset~\(\downarrow\) sweeps, displaying the noise characteristics in different operational states. (g) Schematic representation of the switching mechanism in Al/ZIF-8/ITO device in the pristine state with zero applied voltage to filament formation due to applied voltage to filament formation during SET process to filament rupture during RESET process.}
\label{fig:Figure8}
\end{figure*}

Impedance spectroscopy was employed to further investigate the switching mechanism of the ZIF-8-based resistive switching memory device in both the HRS and LRS\cite{Kotova2018}. Measurements were conducted over a frequency range of 500 Hz to 1 MHz with a DC bias of 200 mV. The Cole-Cole plots (Imaginary Z vs Real Z) for HRS and LRS are illustrated in Figure \ref{fig:Figure4} (e) and Figure \ref{fig:Figure4} (f), respectively. In Figure \ref{fig:Figure4} (e), the Cole-Cole plot for HRS shows a characteristic semi-circular curve, indicative of a parallel resistor-capacitor (RC) network, highlighting capacitive behavior in the high resistance state.\cite{Jonscher1983} Conversely, Figure \ref{fig:Figure4} (f) presents the Cole-Cole plot for the LRS, showing a real impedance within the range of 400 to 900 ohms, with minimal imaginary impedance over varying frequencies. This response suggests a primarily resistive component, where the observed resistance variations reflect fluctuations in the resistive characteristics. The contrasting behavior of the HRS and LRS indicates capacitive characteristics in the HRS and resistive characteristics in the LRS. This impedance behavior strongly implies a filamentary switching mechanism.\cite{Jiang2013} In the LRS, a conductive filament forms between the electrodes, likely due to the migration of metallic ions within the ZIF-8 matrix, with the fluctuating resistance in this filament attributed to the diffusive movement of ions. 

To further confirm the conduction mechanism, low-frequency noise measurements are performed. The time-dependent fluctuations can be analyzed in both the time and frequency domains to reveal distinct conduction mechanisms. Time-domain analysis uncovers stable states through probability distribution functions (PDF) and time-lag plots (TLP). Residual current fluctuations at various bias voltages in the high-resistance state (HRS) are shown in Figure \ref{fig:6} (a). At \(V = 0.1 \, \text{V}\), the system exhibits uniform residual current fluctuations, indicating homogeneous conduction in the ohmic region, as depicted in Figure \ref{fig:6} a(i). As the voltage increases to \(V = 0.6 \, \text{V}\), the system begins to show random telegraphic signals (RTS), as shown in Figure \ref{fig:6} a(ii), which are characteristic of switching events. This RTS behavior is associated with the switching of conduction states due to the presence of charge traps, with the noise becoming more pronounced at higher voltages. At \(V = 1.1 \, \text{V}\), the fluctuations become even more pronounced, signaling a transition to space-charge limited conduction (SCLC), as seen in Figure \ref{fig:6} a(iii). Further increases in voltage reduce the noise, but the fluctuations continue to indicate more complex behavior due to multiple conduction paths, leading to additional complexity in the system’s noise characteristics.

The PDF is presented in Figure \ref{fig:6} (b) for various voltages: at \(V = 0.1 \, \text{V}\), the PDF (Figure \ref{fig:6} b(i)) shows a Gaussian distribution, reflecting the uniform, stable fluctuations typical of ohmic conduction. As the voltage increases to \(V = 0.6 \, \text{V}\), the distribution deviates from Gaussian (Figure \ref{fig:6} b(ii)), indicating the presence of two distinct stable states, consistent with the appearance of RTS. This suggests the system undergoes a switching process between two states as seen in (Figure \ref{fig:6} a(ii)). At \(V = 1.1 \, \text{V}\), the PDF (Figure \ref{fig:6} b(iii)) exhibits a higher peak in the high-current region, which corresponds to the transition to SCLC. The shifting of the peak in the distribution reflects the increasing dominance of one conduction mechanism over another. As voltage increases further, the PDF becomes more complex, showing multiple peaks which are indicative of the involvement of multiple conduction paths.
\begin{figure*}[t]
\centering
\includegraphics[width=\textwidth]{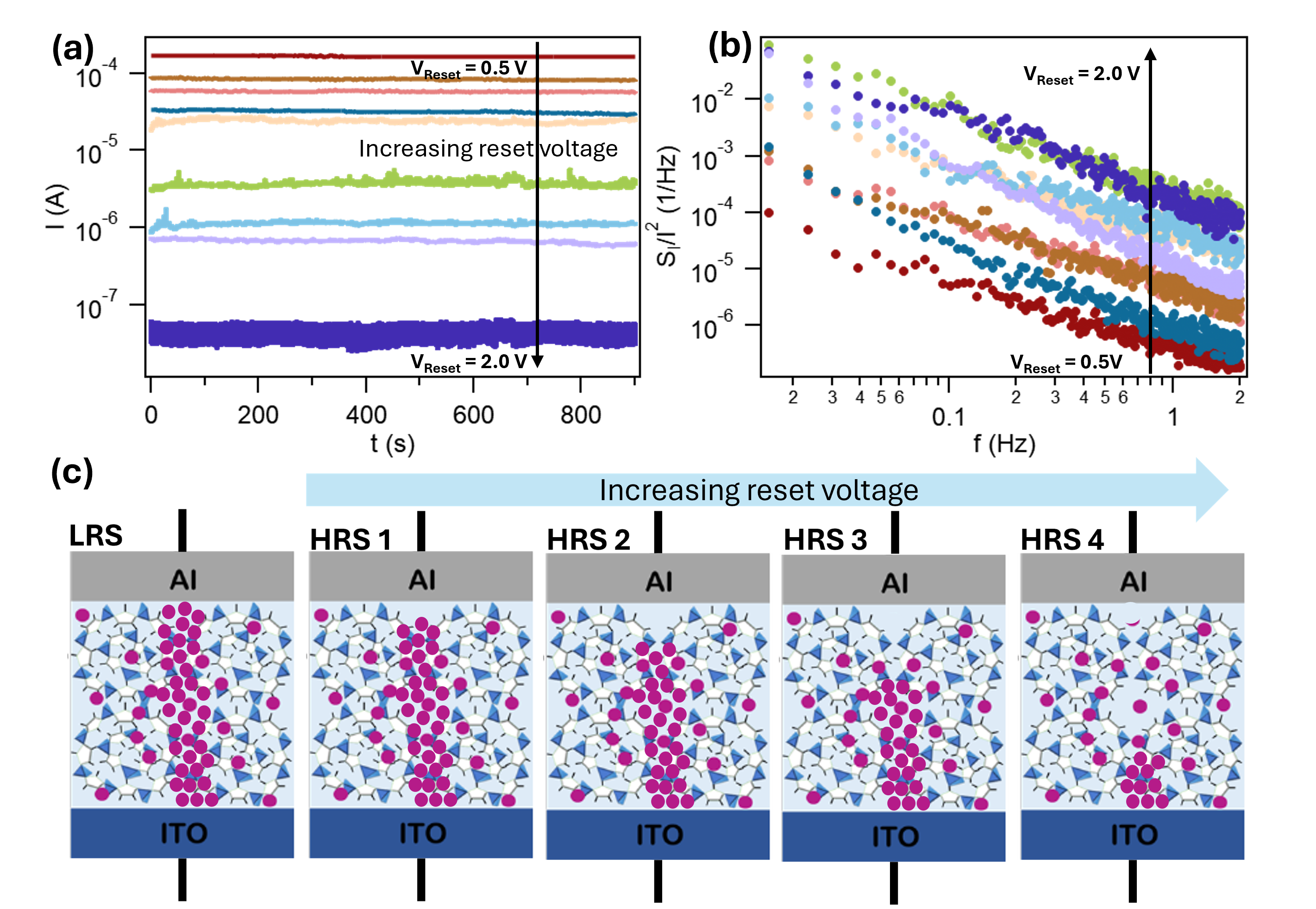}
\caption{Multi-state resistive switching and noise behavior in the device. (a) Time series of current measured at 0.1~V for five distinct states (State~0 to State~4) achieved by applying different reset voltages. The current for each state remains stable for at least 16 minutes. (b) Power spectral density (PSD) of the current fluctuations for each state, showing a decrease in noise magnitude as the reset voltage increases. (c) Schematic of the conductive filament evolution with increasing reset voltage. The filaments progressively shorten as higher reset voltages are applied, transitioning the system through various high resistance states (HRS~1 to HRS~4), each corresponding to a distinct state of conductance.
}
\label{fig:9}
\end{figure*}
The time-lag plots (TLP) in Figure \ref{fig:6} (c) provide insights into the correlations of current fluctuations at different voltages. At \(V = 0.1 \, \text{V}\), the TLP (Figure \ref{fig:6} c(i)) reveals a single, uniform cluster, indicating no significant correlations between fluctuations, as expected in the ohmic region. As the voltage increases to \(V = 0.6 \, \text{V}\), the TLP (Figure \ref{fig:6} c(ii)) displays two diagonal clusters, which are indicative of two stable states in the system. The presence of cross-diagonal clusters reflects transition states, highlighting the switching dynamics of the system. At \(V = 1.1 \, \text{V}\), the TLP (Figure \ref{fig:6} c(iii)) shows an increased density of cross-diagonal points, which corresponds to more frequent transitions between states due to the involvement of multiple conduction paths, consistent with the behavior seen in the SCLC region. Further increases in voltage result in multiple diagonal clusters, suggesting that the system is capable of supporting multiple stable conduction states, likely corresponding to the formation of several filaments.\cite{Das2021} Subsequent sections of the manuscript explore the stability and accessibility of these states, highlighting their potential for multi-level memory applications.

Additionally, frequency-domain analysis offers insights into charge carrier dynamics by examining the magnitude of the PSD and the slope \(\alpha\). The PSD of the current fluctuations, representing the average of the squared current per unit frequency interval, is described by
\begin{equation}
S_I(f) = 2 \lim_{T \to \infty} \frac{1}{T} \left| \int_{-T/2}^{+T/2} dt \, e^{-i\omega t} \delta I(t) \right|^2
\end{equation}
Here, \(\delta I(t)\) represents the time-dependent fluctuation in the current signal, recorded over the time interval \(T\), and \(\omega = 2\pi f\) corresponds to the angular frequency. The PSD is computed using Welch’s periodogram method.\cite{Welch} Noise measurements were conducted throughout the set and reset cycles to comprehensively investigate the device's behavior, as illustrated in Figure~\ref{fig:Figure8}. Figure~\ref{fig:Figure8} (a) showcases four representative noise spectra corresponding to distinct conduction regions: ohmic, SE, and SCLC. These regions exhibit noticeably different PSD magnitudes and slopes. In the ohmic region, the PSD slope is close to 1, reflecting uniform conduction with minimal fluctuations. By contrast, the SE and SCLC regions display significantly steeper slopes, exceeding 1.5. Furthermore, the PSD magnitudes vary substantially, with the SCLC region showing an order-of-magnitude higher noise level compared to SE, as shown in Figure~\ref{fig:Figure8} (c). These distinctions emphasize the unique noise characteristics associated with different conduction mechanisms. During the set cycle, the device initially remains in the high-resistance state (HRS) with stable noise levels at a bias voltage of 0.1 V, as depicted in Figure~\ref{fig:Figure8} (b). As the voltage increases, the migration of Zn ions triggers the formation of conductive filaments, resulting in a rise in noise. This noise reaches its peak when the filament is fully formed. Subsequently, as multiple conductive paths are established, the noise magnitude decreases dramatically—by nearly four orders of magnitude, as illustrated in Figure~\ref{fig:Figure8} (d). Once the filaments are formed, they exhibit non-volatile behavior, ensuring structural stability and unchanged noise magnitudes during the set-down cycle, as shown in Figure~\ref{fig:Figure8} (e). This indicates stable conduction, even under varying voltages. The reset cycle begins with the application of a negative bias, initiating the dissolution of the filaments and returning the device to the HRS. At -0.1~V, the noise magnitude is approximately four orders higher than that observed at +0.1~V, as evident in Figure~\ref{fig:Figure8} (f). This behavior reflects the onset of Zn ion migration and filament breakdown. As the process advances, the noise magnitude reduces by nearly three orders of magnitude, eventually stabilizing. Notably, during the reset-down cycle, the noise magnitudes remain constant, reflecting stable conduction after the complete dissolution of filaments. Throughout the reset cycle, the noise spectra reveal multiple intermediate states before the device fully transitions back to the HRS.

Based on this study, the conduction mechanism in the ZIF-8 matrix, where zinc ions (Zn\textsuperscript{2+}) are coordinated with organic imidazole chains, can be understood as being driven by the oxidation and reduction of these zinc ions, which induce the resistive switching (RS) behavior in the devices.\cite{Xu2017,Jagannadham2022} Under an applied electric field, Zn\textsuperscript{2+} ions delocalize or migrate through the ZIF-8 layer toward the ITO electrode, forming a conductive filament that bridges the Al and ITO electrodes. Additionally, the departure of Zn ions from the ZIF-8 framework leaves unbonded nitrogen atoms in the imidazolate linkers, providing another pathway for electron conduction.\cite{Park2017}  Figure~\ref{fig:Figure8} (g) illustrates the device in its initial high-resistance state (HRS), where no filament is present. Upon applying a positive bias to the Al electrode, Zn\textsuperscript{2+} ions migrate toward the ITO electrode and reduce to Zn atoms. With further accumulation, these Zn atoms, along with the unbonded nitrogen sites left by displaced Zn ions, form a conductive filament that connects the electrodes, shifting the device to a low-resistance state (LRS). When the bias is reversed, the Zn\textsuperscript{2+} ions disperse, breaking the filament and returning the device to HRS. The stable Zn\textsuperscript{2+}-based filament in the LRS enhances retention and endurance. These intermediate states of different filament structure, exhibit distinct resistances, making them viable for multi-bit, high-density memory applications. This demonstrates the significant potential of these devices for advanced memory technologies.

Interestingly, the ZIF-8 device has the potential to reset to multiple resistance states without crossover, enabling the storage of multibit data. As various reset voltages are applied, the current from the device gradually decreases as it transitions to higher resistance states. Figure~\ref{fig:9} (a) illustrates the retention characteristics of the device in each stable, accessible state. The device can maintain 9 distinct states for at least 16 minutes with minimal change. This provides ample margin for utilizing the device to store up to three bits (8 states) per cell. These multiple states arise because different reset voltages break the filament to varying extents, resulting in different HRS. To understand the physical mechanism behind these stable multi-states, noise spectroscopy was performed at different states by applying 0.1~V (the same as the read voltage). The PSD magnitude for these states is shown in Figure~\ref{fig:9} (b). The PSD magnitude increase as higher reset voltages are applied. As the reset voltage increases, the magnitude of PSD increases due to a reduction in the filament length between the top and bottom electrodes, which leads to an increase in \( L \). Figure~\ref{fig:9} (c) schematically illustrates the variation in filament length as observed through noise spectroscopy, highlighting how the filament's structural changes depend on the applied reset voltage and the resulting resistance states.

\section{Conclusion}

In summary, ZIF-8 was successfully synthesized via a simple solution process method and thoroughly characterized for its structural and optical properties. The Al/ZIF-8/ITO memristive device, fabricated using a spin-coating technique, exhibited consistent bipolar resistive switching behavior with set/reset voltages of 1.5V and -1.5V, respectively, over 60 switching cycles. The device demonstrated a high On/Off resistance ratio of $\sim$10$^2$, with excellent data retention of up to 10$^4$ seconds and stable performance. Impedance spectroscopy analysis revealed capacitive behavior in the HRS and pure resistive behavior in the LRS, suggesting a filamentary switching mechanism. Noise spectroscopy revealed same switching mechanism, also provide insights for possible stable multiple states. The migration and subsequent accumulation of Zn$^{+2}$ ions, coordinated with imidazole ligands in the ZIF-8 matrix, under an external bias were identified as the key factors for filament formation. Upon the application of reverse bias, the diffusion of ions caused the rupture of the filament, giving rise to the observed resistive switching behavior. This Zn-based filamentary mechanism imparts promising RS characteristics to the device. Reproducibility and variability were confirmed through the analysis of seven different devices, all of which exhibited consistent performance. Furthermore, the device maintained stability across a wide temperature range, from -20°C to 100°C. The notable RS parameters, such as the high On/Off resistance ratio, excellent retention, and high endurance, underscore the potential of ZIF-8 as a viable candidate for non-volatile ReRAM applications. Moreover, the low-temperature solution-processed approach presents ZIF-8 as a suitable material for integration into future flexible device technologies.

\medskip
\textbf{Supporting Information}
Supporting Information is available from the Wiley Online Library or from the author.

\begin{acknowledgements}

DK and NK contributed equally to this work. DK acknowledges CSIR (Government of India) for grant of SRF. NK acknowledges support from the College of Arts and Sciences at the University at Buffalo. Electrical transport measurements at the University at Buffalo were supported by the National Science Foundation grant NSF-MRI 1726303.
\end{acknowledgements}

\appendix
\section{Supplementary Information}

\begin{figure}[h!]
\centering
\renewcommand{\thefigure}{S1} 
\includegraphics[width=0.5\textwidth]{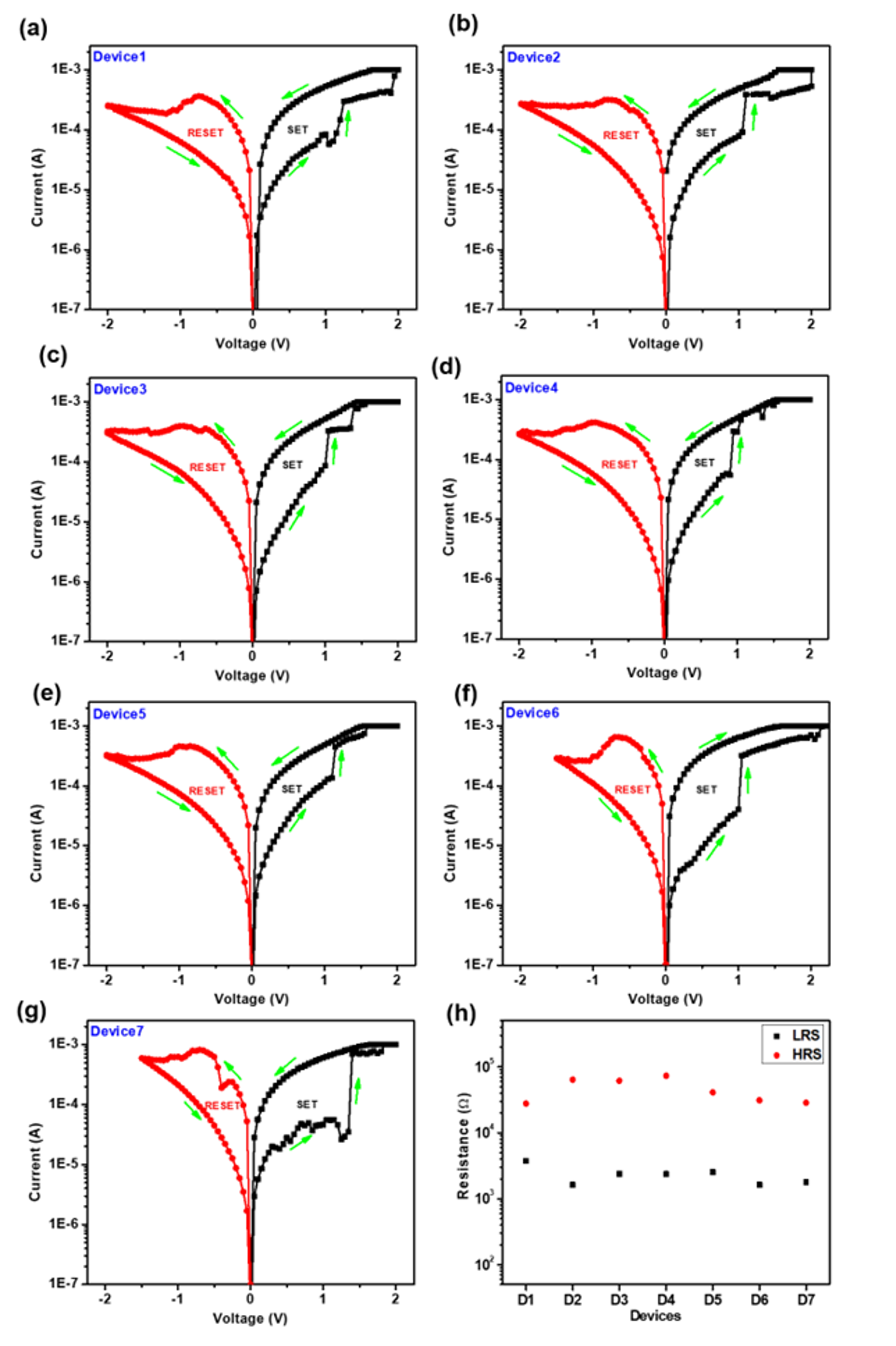}
\caption{(a)-(g) RS behaviour of 7 different Al/ZIF-8/ITO structured devices (h) corresponding HRS and HRS resistances of these devices.
}
\label{fig:S1}
\end{figure}

\begin{figure*}
\centering
\renewcommand{\thefigure}{S2} 

\includegraphics[width=0.8\textwidth]{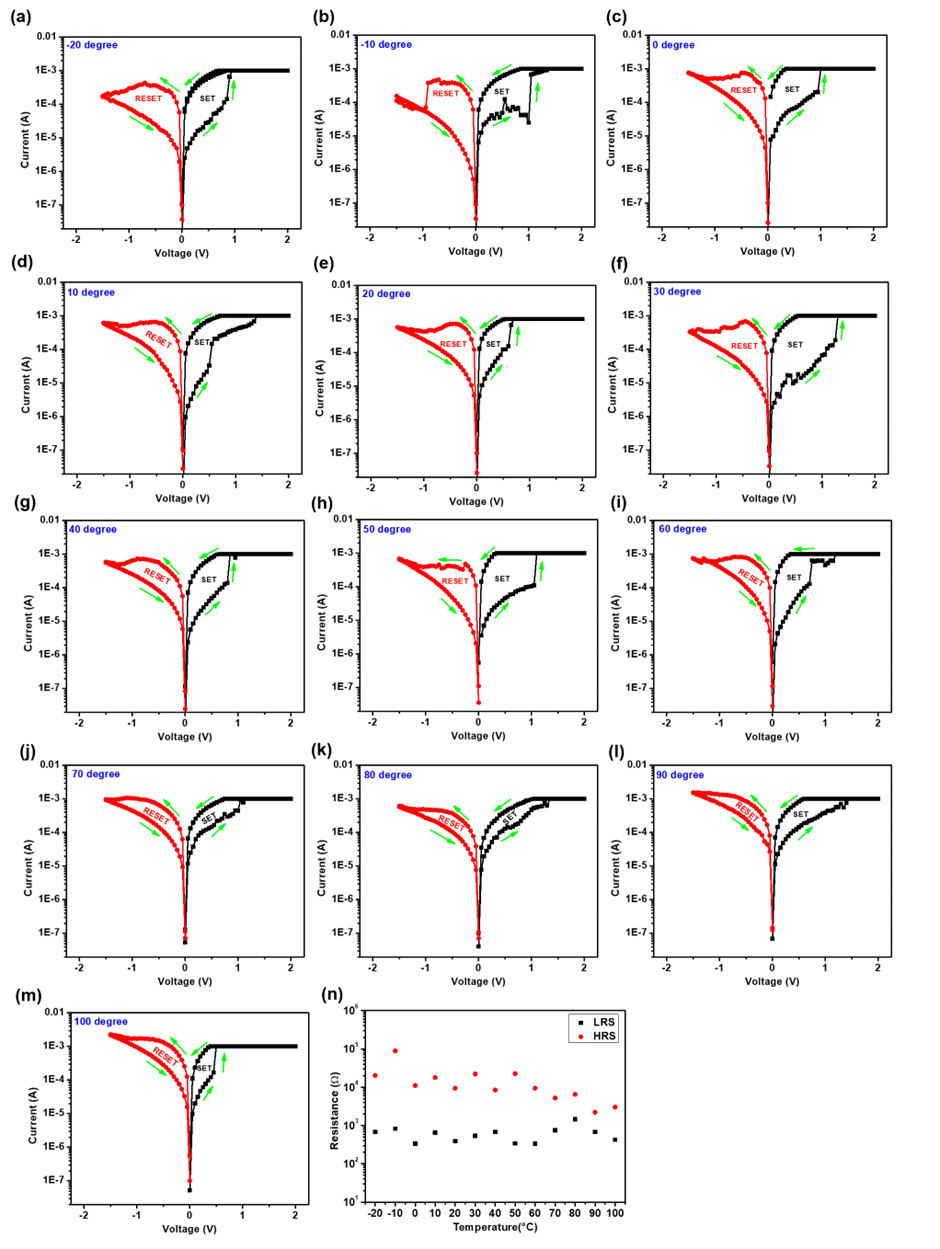}
\caption{(a)-(m) RS behaviour of Al/ZIF-8/ITO structured devices at different temperature ranging between -20°C to 100°C, and (n) corresponding HRS and HRS resistances of the device at these temperatures.
}
\label{fig:S2}
\end{figure*}

\clearpage
\bibliographystyle{MSP}
\bibliography{Ref}

\end{document}